\documentclass[a4paper,11pt]{article}
\pdfoutput=1
\usepackage{jheppub}

\usepackage{hyperref}
\usepackage{amsmath,amssymb,amsfonts,amsthm}
\usepackage{graphicx}
\usepackage{enumitem}
\usepackage{bm}
\usepackage{rotating}
\usepackage{pbox}
\usepackage{array}
\usepackage{mathtools}
\usepackage{leftidx}
\usepackage{lmodern}
\usepackage[normalem]{ulem}
\usepackage{braket}
\usepackage{tensor}
\usepackage[usenames,dvipsnames]{xcolor}
\usepackage{pifont}
\usepackage{float}
\usepackage{footnote}
\usepackage{setspace}
\usepackage{dsfont}

\newcommand{\openone}{\mathds{1}}

\newcommand{\R}{\mathbb{R}}
\renewcommand{\d}{\mathrm{d}}
\newcommand{\dd}{\mathrm{d}}
\newcommand{\rr}[1]{\left(#1\right)}

\newcommand{\sx}{{\mathsf{x}}}

\newcommand{\ii}{\text{i}}
\newcommand{\W}{\text{W}}
\DeclareMathOperator{\tr}{\text{Tr}}
\renewcommand{\tilde}{\widetilde}

\newcommand{\osc}{\text{osc}}
\newcommand{\zm}{\text{zm}}
\renewcommand{\bar}{\overline}

\newcommand{\ketbra}[2]{\ket{#1}\!\!\bra{#2}}
\newcommand{\bk}{\bm{k}}
\newcommand{\A}{\mathcal{A}}

\begin{document}

\title{Particle detectors under chronological hazard}

\author[a,b]{Ana Alonso-Serrano}
\author[c,d,e]{Erickson Tjoa}
\author[f]{Luis J. Garay}
\author[g,e,h]{Eduardo~Mart\'in-Mart\'inez}

\affiliation[a]{Institut für Physik, Humboldt-Universität zu Berlin, Zum Großen Windkanal 6, 12489 Berlin, Germany}
\affiliation[b]{Max-Planck-Institut f\"ur Gravitationsphysik 	(Albert-Einstein-Institut), Am M\"uhlenberg 1, D-14476 Golm, Germany}
\affiliation[c]{Max-Planck-Institut f\"ur Quantenoptik, Hans-Kopfermann-Stra\ss e 1, D-85748 Garching, Germany}
\affiliation[d]{Department of Physics and Astronomy, University of Waterloo, Waterloo, Ontario, N2L 3G1, Canada}
\affiliation[e]{Institute for Quantum Computing, University of Waterloo, Waterloo, Ontario, N2L 3G1, Canada}
\affiliation[f]{Departamento de F\'isica Te\'orica and IPARCOS, Universidad Complutense de Madrid, 28040 Madrid, Spain}
\affiliation[g]{Department of Applied Mathematics, University of Waterloo, Waterloo, Ontario, N2L 3G1, Canada}
\affiliation[h]{Perimeter Institute for Theoretical Physics, 31 Caroline St N, Waterloo, Ontario, N2L 2Y5, Canada}

\emailAdd{ana.alonso.serrano@aei.mpg.de}
\emailAdd{erickson.tjoa@mpq.mpg.de}
\emailAdd{luisj.garay@ucm.es}
\emailAdd{emartinmartinez@uwaterloo.ca}

\date{\today}

\abstract{
We analyze how the presence of closed timelike curves (CTCs) characterizing a time machine can be discerned by placing a local particle detector in a region of spacetime which is causally disconnected from the CTCs. Our study shows that not only can the detector tell if there are CTCs, but also that the detector can separate topological from geometrical information and distinguish periodic spacetimes without CTCs (like the Einstein cylinder), curvature, and spacetimes with topological identifications that enable  time-machines.
}

\preprint{IPARCOS-UCM-24-015}
\maketitle

\section{Motivation}

The theoretical underpinnings of time machines are grounded in general relativity, which permits solutions with intricate causal structures like Gödel's rotating universe or wormhole spacetimes. Traversable wormholes, if they exist, offer theoretical pathways to constructing time machines, where closed timelike curves (CTCs) occur within certain regions of spacetime. However, the presence of closed timelike curves typically leads to Cauchy horizons in these spacetimes, with the stability of viable time machines  linked to the stability of these horizons. 

Within the framework of quantum field theory (QFT) in curved spacetimes, the instability of Cauchy horizons can be assessed by analyzing the divergences in the renormalized stress-energy tensor near the Cauchy horizons. Since the background geometry containing a time machine is necessarily multiply-connected and not globally hyperbolic, defining a QFT on such a background spacetime requires the introduction of new tools. Extensive research has been devoted to the general construction of QFT on multiply connected manifolds, often employing the framework of automorphic fields. The approach involves studying the quantum fields on the universal covering space, which has a trivial topology, while imposing specific automorphic conditions on them  \cite{BanachDowker1979,Banach1980automorphic,Frolov1991locallystatic}.  

A classic illustration of the automorphic construction involves a scalar field on the Einstein cylinder. In this scenario, a scalar field on the cylinder is determined by the field configuration in Minkowski space (the universal covering of the cylinder), subject to (anti-)periodic boundary conditions along the spatial direction. For massless scalar fields in flat space subject to periodic or Neumann boundary conditions, or when the spatial sections of curved spacetime are compact, it is known that canonical quantization can be somewhat pathological due to the appearance of zero modes \cite{EMM2014zeromode,tjoa2019zeroresponse,tjoa2020harvesting,Page2012deSitter,Allen1987deSitter}. Since the zero mode is dynamically equivalent to a free particle, it admits no Fock representation and hence the vacuum state of the full theory is ambiguous. A more modern interpretation of this phenomenon is that the zero mode Hilbert space does not have any preferred vacuum state, hence the full theory has a continuum family of unitarily inequivalent vacuum states \cite{witten2021does}. 

Another example, which will be our main object of study in the work, is the time machine spacetimes obtained by suitable topological identification of Anti-de Sitter (AdS) spacetime, which is not globally hyperbolic itself.  QFT on AdS geometry has been studied to great detail by many authors in very different contexts \cite{Isham1978,wald1980nonhyperbolic,Ishibashi2004AdS,Dappiaggi2016PAdS,Dappiaggi2018CAdS}, often due to its current importance for Anti-de Sitter/Conformal Field Theory (AdS/CFT) correspondence and holography (see, e.g., \cite{maldacena1999largeN,Witten:1998zw,de2001holographic}).

In a previous paper \cite{Ana2021timemachine}, it was proposed that one can single out a ``regularized'' vacuum state for the zero mode of the Einstein cylinder by considering a scalar field living on a suitable topological identification of the Poincar\'e patch of two-dimensional Anti-de Sitter (AdS$_2$) spacetime. In this so-called \textit{Poincar\'e time machine}\footnote{Due to conformal flatness in two dimensions, the time machine considered here is locally equivalent to the Misner-AdS$_2$ time machine in \cite{emparan2021holography}.} background, a fixed (large) AdS length serves as a natural regulator for the zero mode.

In this work we will extend the analysis in \cite{Ana2021timemachine} to study the possible extraction of topological information associated with the time-machine model by a observer carrying localized quantum-mechanical probe. In particular, we will see that a localized particle detector  will be able to tell if there are closed-timelike curves in some region of the spacetime far from the detector's trajectory (even though this region is hidden by a horizon). A natural framework for this is the so-called {Unruh-DeWitt} (UDW) particle detector model~\cite{Unruh1979evaporation,DeWitt1979}, which has been argued to be a good model for the light-matter interaction consisting of a localized two-level quantum system interacting with the quantum field via dipole-type interaction in quantum optics \cite{Lopp2021deloc}. 
The localized nature of the interaction enables us to obtain useful local information about the field that is consistent with what local observers can measure. Crucially, in the case of the Einstein cylinder, the choice of zero modes will have phenomenological impact since the UDW detector couples to all the modes of the field including the zero mode  \cite{EMM2014zeromode,tjoa2019zeroresponse,tjoa2020harvesting}. For our purposes, we will use a variation of the UDW detector model called the {derivative coupling model} \cite{Aubry2014derivative,juarez2018quantum,tjoa2022unruhdewitt} since it is less sensitive to the infrared (IR) ambiguities that appear in two-dimensional massless QFT.

Note that the goal of this paper is not  simply to study the dynamics of an UDW detector applied to an AdS background geometry and its topological identifications (this has been studied in, e.g., \cite{Deser1997,Jennings2010,Hodgkinson2012BTZ,Henderson2020antihawking,Pitelli2021ads2}). Furthermore, it is known that measurements made with localized probes can indeed be sensitive to the topology of spacetime and not only its geometry~\cite{smith2016topology}. Armed with this knowledge, we also want to  understand under what conditions a measurement carried out by a localized probe can distinguish between a spacetime with a time machine and spacetimes with similar geometries where there are no time machines even when the detector  does not travel on a CTC itself.

We begin by observing that the Ricci scalar of the time machine geometry considered in \cite{Frolov1991locallystatic,Ana2021timemachine} is constant and determined by the two parameters characteristic of the periodic identification that leads to the time machine: the spatial period of the time-machine construction and the strength of the time-machine, that is, what is the time warp in the identification. Note that already for the zero-curvature limiting case, taking the limit of vanishing time warp and is not equivalent to taking the limit of large spatial periods: while both limits recover (locally) flat spacetime, the former leads to the Einstein cylinder and the latter to Minkowski spacetime. This suggests that for fixed curvature, keeping the spatial period fixed and varying the time-machine strength is not physically equivalent to keeping the latter fixed while varying the former. 
We can thus isolate the global topological and chronological information associated with the existence of the time machine from the geometrical information by fixing the local curvature in two distinct ways\footnote{This is a curved spacetime generalization to the result from \cite{smith2016topology}, where the UDW detector can tell the difference between Minkowski spacetime vs Einstein cylinder that do not share the same topology at a timescale shorter than the light-crossing time of the cylinder.}.  In particular, we will concentrate on the two regimes that we call ``slow'' and ``fast'' time machines with the same AdS curvature\footnote{We borrow the well-chosen terminology of \cite{emparan2021holography} when they considered the time machine geometry of AdS-Misner type.} where only in the second limit the Poincar\'e-AdS$_2$ limit is reached.

In this paper we will first show that the detector is sensitive to  topological information of the background geometry as well as the presence of CTCs beyond the bifurcate Cauchy horizon. We then show that for the Poincar\'e time machine geometry, the two time-machine configurations are physically inequivalent and that this is operationally manifest in terms of very different detector responses in each regime. The fact that local measurements of particle detectors can be used to track non-local features of a QFT in regions that are causally disconnected from the detector is a well studied phenomenon, and can be ultimately traced back to the fact that equilibrium states of quantum fields store global information locally~\cite{KeithCo,smith2016topology,Aida1,KeithCo2,Smith_CQG,Aida2,MachineLearning}.

Our paper is organized as follows. In Section~\ref{sec: qft} we briefly review the scalar field theory in two-dimensional geometries, focusing on the Einstein cylinder and the Poincaré time machine geometry. In Section~\ref{sec: derivative-coupling-model} we review the derivative coupling Unruh-DeWitt model and calculate the relevant derivative-coupling Wightman functions and how they are related in appropriate limits. In Section~\ref{sec: detector-response} we calculate the detector response and show how it can be used to understand the slow and fast time machine regimes.  We adopt the convention that $c=\hbar=1$ and the metric signature is such that for a timelike vector $\mathsf{V}$ we have $g(\mathsf{V},\mathsf{V}) = g_{\mu\nu}V^\mu V^\nu <0$.

\section{Scalar QFT in multiply connected spacetimes}
\label{sec: qft}

For the purpose of this work, we will consider quantum field theory for a massless scalar field in three different two-dimensional spacetimes: (1) the time machine geometry constructed by topological identification of the Poincar\'e patch of anti-de Sitter (AdS$_2$), which we shall call ``Poincar\'e time machine'' for brevity; (2) the universal covering space of the time machine, i.e., the full Poincar\'e patch itself; and (3) the Einstein cylinder, a flat spacetime with topology $\R\times S^1$. Following \cite{Ana2021timemachine}, we will label quantities in the time-machine spacetime with a bar: for example, the field living in Poincar\'e AdS$_2$ is written as $\phi$, while the field on the time machine geometry with $\bar{\phi}$. Quantities associated with the Einstein cylinder will be written in a manner that clearly distinguishes from the ones for the Poincar\'e patch and the time machine, by using capital letters.

A real massless scalar field $\phi$ in  $(1+1)$-dimensional curved spacetime conformally coupled to gravity obeys the Klein-Gordon equation
\begin{align}
    (-\nabla_\mu\nabla^\mu)\phi = 0\,.
    \label{eq: KGE-curved}
\end{align}
Recall that canonical quantization of a real scalar field gives us an operator-valued distribution $\hat\phi(\sx)$ whose mode decomposition is given by
\begin{align}
    \hat\phi(\sx) = \sum_j \hat a_j u_j(\sx) + \hat a_j^\dagger u_j^*(\sx)
\end{align}
where $u_j$ are the eigenmodes of Eq.~\eqref{eq: KGE-curved} and the sum is understood to be over continuous or discrete set of modes depending on the spectrum. In this work we will consider both the time machine geometry (where the basis that we will use and hence $j$ are discrete) and its universal cover, namely the full Poincar\'e patch of AdS$_2$ (where they will be continuous).  The operators $\hat a_j,\hat a_j^\dagger$ are the creation and annihilation operator satisfying the canonical commutation relation
\begin{align}
    [\hat a_j,\hat a_{j'}^\dagger] = \delta_{jj'}\openone\,,\quad [\hat a_j,\hat a_{j'}] = [\hat a_j^\dagger,\hat a_{j'}^\dagger]=0\,,
\end{align}
where $\delta_{jj'}$ represents Kronecker delta if $j$ is discrete or Dirac delta function if $j$ is continuous.

\subsection{QFT in time machine geometry and its universal cover}
\label{subsec: QFT-timemachine}

In this section we summarize the main results obtained in a previous paper by the authors~\cite{Ana2021timemachine} on the computation of Wightman function for fields on a  $(1+1)$  canonical time machine  $\bar M$~\cite{Ana2021timemachine}. The metric on the time machine geometry is given by
\begin{align}
    \d s^2=-e^{-2\W y}\dd t^2+\dd y^2\,,\hspace{0.5cm} \W=\frac{\log A}{L},
    \label{eq:time-machine}
\end{align}
where $A\geq 1$ is the warp parameter giving an estimation of the ``strength'' of the time machine and $L > 0$ is the proper length. In order to  have a  time machine points are identified via the equivalence relation $(t,y) \sim (t',y')$ if and only if $t'/t=A$ and $y'-y=L$, giving rise to a multiply connected spacetime. One can directly see that this model is fully characterized by two parameters  $A$ and $L$. The universal covering space $M$ of this multiply-connected spacetime is the \textit{Poincar\'e patch} of AdS$_2$ spacetime with line element  \eqref{eq:time-machine} (one can see that the canonical time-machine model $\bar M$ is the quotient space obtained by previous identification of the Poincar\'e patch).

We use universal covering techniques \cite{Dowker1972multiplyconnected, Banach1979mathissues, Banach1980automorphic,Frolov1991locallystatic} to construct a field theory on the multiply-connected spacetime. In simple spacetimes such as the Einstein cylinder, characterized by $A=1$,  these techniques reduce to just imposing boundary conditions on the field, and in our $(1+1)$-time-machine model the procedure involves constructing the automorphic field $\bar\phi$ from the corresponding field $\phi$ living on the Poincar\'e patch of AdS$_2$ (for more details on the construction of the model, see~\cite{Ana2021timemachine}).

Let us now describe a change of variables to the null-coordinates that we will use in this paper. We start by writing the metric in the more standard Poincaré-patch coordinates $(\eta,\xi)\in \mathbb R\times\mathbb R_+$ by the change of coordinates
\begin{align}
    \eta = t\,,\hspace{0.5cm} \xi =  e^{\W y}/\W\,.
    \label{eq:AdS-metric}
\end{align}
The metric takes the form
\begin{align}
    \dd s^2 = \frac{1}{\W^2\xi^2}\rr{-\dd \eta^2+\dd \xi^2}\,,
    \label{eq:poincarepatch0}
\end{align}
where the AdS$_2$ length scale is given by $\W^{-1}$. The null-coordinates are then defined by
\begin{align}
    \zeta_\pm = \xi \pm \eta\,,
\end{align}
in terms of which the topological identification that gives rise to the time-machine model reads $(\zeta_+,\zeta_-)\sim A(\zeta_+,\zeta_-)$.

\begin{figure}[tp]
    \centering
    \includegraphics[scale=0.8]{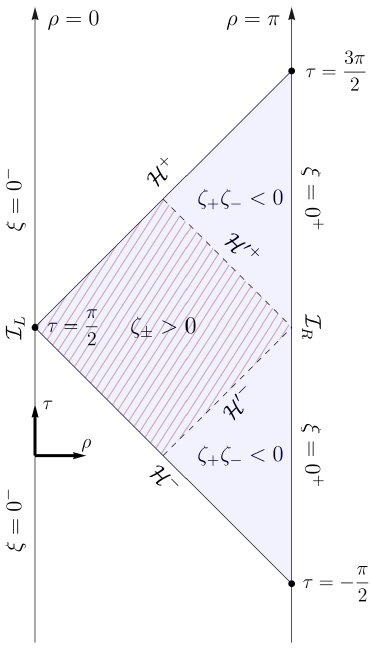}
    \caption{Conformal diagram for the Poincar\'e patch of AdS$_2$. The  Cauchy horizons  of the Poincar\'e patch are labelled $\mathcal{H}^\pm$.  After topological identification, the region $\zeta_+\zeta_-<0$ presents CTCs, thus two new Cauchy horizons $\mathcal{H'}^\pm$ appear in the conformal diagram.  In the compactified coordinates $(\tau,\rho)$, the Poincar\'e patch is covered by $\rho>|\tau-\pi/2|$ and $\rho\in (0,\pi)$.}
    \label{fig: penrose}
\end{figure}

Finally, in order to clarify the structure of regions in the model, we display the Penrose diagram of the maximal analytic extension of the Poincaré patch. For that purpose we first   transform to yet another new set of coordinates $(\tau,\rho)$ defined by
\begin{equation}
    \tan(\rho\pm\tau)=2\W\zeta_\pm.
\end{equation}
Thus, the Penrose diagram of the maximal analytic extension (see Figure \ref{fig: penrose}) is defined by the range $\rho\in(0,\pi), \tau\in \mathbb R$ with a conformal boundary $\mathcal{I}$, consisting of the disconnected $\mathcal{I}_{L}$ at $\rho=0$ and $\mathcal{I}_R$ at $\rho=\pi$. The Poincaré patch, our covering space $M$, covers the  colored region $\rho>|\tau-\pi/2|$ and it is bounded by two past and future Cauchy horizons at $\rho=|\tau-\pi/2|$, i.e. at \mbox{$\zeta_+=\infty$}  and  $\zeta_-=\infty$, respectively. Beyond the horizons AdS spacetime possesses CTCs\footnote{However, working only with AdS would not allow us to decouple the topology of the spacetime from its geometry in order to identify their contributions to detectors response independently.}. The time-machine model $\bar M$ introduces new Cauchy horizons  $\mathcal{H'}^\pm$ defining new regions beyond them with CTCs, such that only the diamond-shaped region $\zeta_\pm>0$ is free of CTCs.

In order to construct our field theory, we perform now the quantization via the universal covering approach to the standard Poincar\'e patch. There, the massless Klein-Gordon equation is $(-\partial_\eta^2+\partial_\xi^2)\phi=0$ with Dirichlet boundary conditions, i.e. $\phi|_{\xi=0}=0$~\cite{Ana2021timemachine}  at the conformal boundary $\xi=0$, as AdS$_2$ is not globally hyperbolic.

We obtain a canonical quantization of the field $\phi$ in terms of $\zeta_\pm$ coordinates that can be written as
\begin{equation}
    \phi(\zeta_+,\zeta_-)=\int_0^\infty  \!\!\dd \omega  \bigr[a(\omega) u_\omega(\zeta_+,\zeta_-)+a(\omega)^\dagger u_\omega^*(\zeta_+,\zeta_-)\bigr],
\end{equation}
where  the positive-frequency eigenfunctions are given by
\begin{align}
    u_\omega(\zeta_+,\zeta_-) = \frac{1}{\sqrt{4\pi\omega}}(e^{-\ii\omega \zeta_+}-e^{\ii\omega \zeta_-})\,,
    \label{eq: positive-frequency-eigenmode}
\end{align}
and $a(\omega)$ and $a^\dagger(\omega)$ are the annihilation and creation operators acting on the Fock space defined in terms of the positive frequency modes, with commutation relation $[a(\omega) ,a ({\omega'})^\dagger]=\ii\delta(\omega-\omega')$.

We can obtain the corresponding QFT defined on the time-machine spacetime $\bar M$ from the QFT defined in the covering space $M$. We need to construct $\bar\phi(\zeta_+,\zeta_-)$, which is defined in the fundamental domain $\W(\zeta_+,\zeta_-)\in (1,A)$, from $\phi(\zeta_+,\zeta_-)$, which is defined in the universal cover. The topological identification implies that the values of the scalar fields in the time-machine spacetime have to coincide in both identified points, $\bar\phi(A\zeta_+,A\zeta_-)=\bar\phi(\zeta_+,\zeta_-)$. That is, that the field $\bar\phi$ has to be automorphic under the action of the fundamental group, and it is imposed by means of the following requirement in the ``annihilation variable'' $\bar a(\omega)$

\begin{align}
    \bar a(\omega)=\sqrt A\ \bar a(A\omega)\,,
    \label{eq: automorphic-ladder}
\end{align}
that can be satisfied if $\bar a(\omega)$ takes the following form~\cite{Ana2021timemachine}
\begin{align}
    \bar a(\omega) =\omega^{-1/2} \sum_{n=-\infty}^\infty \bar c_n (\omega/\W)^{-2\pi \ii n \beta}\,,\hspace{0.25cm}\beta = \frac{1}{\log A}\,,
    \label{eq: Fourier-hidden-auto}
\end{align}
where $\bar c_n$ are constants fully determined by $a(\omega)$ and vice versa. 

We can write the decomposition of the automorphic solutions $\bar\phi(\zeta_+,\zeta_-)$ in $\bar{M}$ as the infinite sum 
\begin{align}
    \bar\phi(\zeta_+,\zeta_-) = \sum_{n=-\infty}^\infty [\bar c_n \bar{u}_n(\zeta_+,\zeta_-)+\bar c_n^\dagger \bar{u}^*_n(\zeta_+,\zeta_-)]\,,
\end{align}
where $\bar c_n^{\phantom{*}}$, $\bar c_n^\dagger$ are annihilation and creation operators acting  on the Fock space defined in terms of the positive frequency modes $\bar u_n$, with canonical commutation relations $[\bar c_n^{\phantom{\dag}},\bar c_{n'}^\dag]=\ii \delta_{nn'}$. The normalized modes $\bar{u}_n$ under the induced Klein-Gordon inner product in the fundamental domain~\cite{Ana2021timemachine} are given by
\begin{align}
    \overline{u}_0(\zeta_+,\zeta_-) &=- \rr{\frac{\beta}{4\pi}}^{\frac{1}{2}}\bigg(\ln\frac{|\zeta_+| }{|\zeta_-| }+\ii\frac{\pi}{2}(s_++s_-)\bigg)\,,
    \label{eq: u0}\\
    \overline{u}_{n\neq0}(\zeta_+,\zeta_-) &=[8\pi n \sinh(2\pi^2 \beta n)]^{-\frac{1}{2}}\notag\\
    &\hspace*{-2em}\times\bigr(e^{-\pi^2\beta n s_+}|\W\zeta_+| ^{2\pi \ii \beta n  }  -e^{\pi^2\beta n s_-}|\W\zeta_-|^{2\pi \ii \beta n }\bigr)\,,
    \label{eq: uj}
\end{align}
where $s_\pm=\text{sign}(\zeta_\pm)$, and the positive frequency modes $\bar{u}_n$ form an orthonormal basis $(\bar u_n, \bar u_{n'})=\delta_{nn'}$. 

Let us remark that for the rest of the paper we perform the computations in the region where there are no CTCs in the time-machine spacetime. As we mentioned before, this corresponds to the  diamond-shaped region $\zeta_\pm>0$ of the spacetime diagram(for more details, see the discussion in~\cite{Ana2021timemachine}).

\subsection{(1+1) Einstein cylinder}
\label{subsec: einstein-cylinder}

The Einstein cylinder is obtained from two-dimensional Minkowski spacetime with flat line element
\begin{align}
    {\d s}^2 = -\d t^2+ \d y^2\,
\end{align}
by a topological identification $(t,y)\sim (t,y+L)$, where $L$ is the circumference of the cylinder and $t\in \mathbb R$. The Klein-Gordon equation \eqref{eq: KGE-curved} for the scalar field $\Phi$ is that of flat-space wave equation with periodic boundary condition
\begin{align}
   (-\partial_t^2+\partial_y^2)
   \Phi = 0\,,\hspace{0.5cm}\Phi(t,y) = \Phi(t,y+L)\,.
\end{align} The resulting massless quantum scalar field has Fourier mode decomposition given by \cite{EMM2014zeromode,tjoa2019zeroresponse}
\begin{align}
    \Phi(t,y) &= Q_\zm(t) + \Phi_\osc(t,y)\,,\\
    \Phi_\osc(t,y) &= \sum_{n\neq 0} \frac{1}{\sqrt{4\pi|n|}}\rr{ b_n^{\phantom{\dagger}} e^{-\ii|k_n|t+\ii k_n y}+ \text{h.c.}}\,,
    \label{eq: Fourier-EC}
\end{align}
where $k_n = 2\pi n/L$ and $n\in \mathbb{Z}\setminus\{0\}$. We call $\Phi_\osc$ the \textit{oscillator modes} and the spatially constant piece $Q_\zm(t)$ the \textit{zero mode} because it corresponds to a zero-frequency oscillator. The ladder operators $b_n^{\phantom{\dagger}}, b^\dagger_n$ satisfy the canonical commutation relation $[b_m^{\phantom{\dagger}},b_n^\dagger] = \delta_{mn}$ for all $n,m\neq 0$.

A massless scalar field on Einstein cylinder has a non-unique ground state due to the existence of the zero mode \cite{EMM2014zeromode,tjoa2019zeroresponse}. We can define the Fock vacuum $\ket{0_\osc}$ for $\Phi_\osc$ as the state satisfying $b_n\ket{ 0_\osc}=0$ for all $n\neq 0$, but the zero mode is dynamically equivalent to a free particle and hence does not have a Fock ground state. The zero mode is naturally associated with position and momentum operators $ Q^\textsc{s}_{\zm}, P^\textsc{s}_{\zm}$ respectively (the subscript ``S'' denotes the Schr\"odinger picture) which obey equal-time canonical commutation relation $[ Q^\textsc{s}_{\zm}, P^\textsc{s}_{\zm}] = \ii \openone $. We can write the zero mode $\bar Q_\zm(t)$ as 
\begin{align}
     Q_\zm(t) = Q^\textsc{s}_{\zm} + \frac{ P^\textsc{s}_{\zm}t}{L}\,.
    \label{eq: zero-mode-operator}
\end{align}
In \cite{Ana2021timemachine}, a way to prescribe a (regularized) zero mode ground state was given by taking the zero-curvature expansion of the quantum scalar field on the time machine geometry.

\section{Particle detector model}
\label{sec: derivative-coupling-model}

In this section we calculate within second-order perturbation theory the detector response in the time machine geometry covering the ``slow time machine'' and ``fast time machine'' regimes. We will then compare the results with the detector responses in the Einstein cylinder and Poincar\'e-AdS$_2$ geometries. In order to avoid certain IR issues due to the zero mode, we will use a variant of the Unruh-DeWitt detector model known as the {derivative coupling model} (see, e.g,  \cite{Aubry2014derivative,juarez2018quantum,tjoa2022unruhdewitt}).

\subsection{The setup}

The derivative coupling variant of the Unruh-DeWitt model  is defined by the interaction Hamiltonian \cite{Aubry2014derivative,juarez2018quantum,tjoa2022unruhdewitt}
\begin{align}
    \hat{H}_I(\tau) &= \lambda\chi(\tau)\hat\mu(\tau)\otimes \partial_\tau\hat\phi(\sx(\tau))\,.
    \label{eq: derivative-coupling-Hamiltonian}
\end{align}
The interaction occurs along the detector's trajectory $\sx(\tau)$ parametrized by its proper time $\tau$. That is, the detector's monopole operator $\hat\mu(\tau)$ is linearly coupled to the \textit{proper-time derivative} of the field. Here $\lambda$ is the coupling constant, $\chi(\tau)$ is the switching function controlling the duration of interaction, and $\hat\mu(\tau)$ is the monopole moment of the detector, given by
\begin{align}
    \hat\mu(\tau) = \hat \sigma_+e^{\ii\Omega \tau}+\hat \sigma_- e^{-\ii\Omega \tau}\,.
\end{align}
The operators $\hat\sigma_\pm$ are the $\mathfrak{su}(2)$ ladder operators defined by $\hat \sigma_+\ket{g}=\ket{e}$ and $\hat\sigma_-\ket{e}=\ket{g}$,  where $\{\ket{g},\ket{e}\}$ are the ground and excited states of the detector's free Hamiltonian $\hat H_0 = \Omega\hat \sigma_+\hat \sigma_-$, and $\Omega$ is the energy gap. 

The time evolution operator is given by the unitary
\begin{align}
    \hat U = \mathcal{T}\exp\left[-\ii\int_{-\infty}^\infty\!\!\dd\tau\,\hat H_I(\tau)\right]\,,
\end{align}
where $\mathcal{T}$ is the time-order operator. Working to second order in perturbation theory, the time evolution can be expanded in a Dyson series
\begin{subequations}
\begin{align}
    \hat U &= \openone + \hat U^{(1)} + \hat{U}^{(2)} + \mathcal{O}(\lambda^3)\,,\\ 
    \hat U^{(1)} &= -\ii\int_{-\infty}^\infty\!\!\!\dd\tau\,\hat H_I(\tau)\,,\\
    \hat U^{(2)} &= -\int_{-\infty}^\infty\!\!\!\dd\tau\int_{-\infty}^\infty\!\!\!\dd\tau'\,\Theta(\tau-\tau')\hat H_I(\tau)\hat H_I(\tau')\,,
\end{align}
\end{subequations}
where $\Theta(z)$ is the Heaviside function. Suppose that the initial state of the joint detector-field system is uncorrelated, i.e., 
\begin{align}
    \hat{\rho}_0 &= \hat{\rho}_{\textsc{d},0}\otimes\hat{\rho}_{\phi,0}\,.
\end{align}
The final state of the detector can then be computed perturbatively as
\begin{align}
    \hat{\rho}_\textsc{d} &= \hat\rho_{\textsc{d},0} + \hat{\rho}^{(1)}_{\textsc{d}}+ \hat{\rho}^{(2)}_{\textsc{d}} + \mathcal{O}(\lambda^3)\,,
\end{align}
where the correction term $\hat{\rho}^{(j)}_{\textsc{d}}$ is  $\mathcal{O}(\lambda^j)$:
\begin{equation}
    \begin{aligned}
    \hat{\rho}^{(1)}_{\textsc{d}} &= \tr_\phi\rr{\hat U^{(1)}\hat\rho_{0} + \hat\rho_{0}\hat U^{(1)\dagger}},
    \\
    \hat{\rho}^{(2)}_{\textsc{d}} &= \tr_\phi\rr{\hat U^{(1)}\hat\rho_{0}\hat U^{(1)\dagger}+ \hat U^{(2)}\hat\rho_{0}+\hat\rho_{0}\hat U^{(2)\dagger }}.
\end{aligned}
\end{equation}
For our purposes, we are interested in the case where the field is in some vacuum state, namely $\hat{\rho}_{\phi,0} = \ketbra{0}{0}$. This state has vanishing odd-point functions, hence $\hat{\rho}_{\textsc{d}}^{(1)}=0$ and the leading-order correction to the detector's density matrix is $\mathcal{O}(\lambda^2)$.

For simplicity, let us assume that the detector's initial state is the ground state $\hat{\rho}_{\textsc{d},0} = \ketbra{g}{g}$. Then in the $\{\ket{g},\ket{e}\}$ basis, we have
\begin{align}
    \hat\rho_{\textsc{d}} &= 
    \begin{pmatrix}
    1-P & 0 \\ 0 & P
    \end{pmatrix} + \mathcal{O}(\lambda^4)\,,
\end{align}
where $P$ is the excitation probability given by
\begin{align}
    P &= \lambda^2\int\dd\tau\,\dd\tau'\,\chi(\tau)\chi(\tau')e^{-\ii\Omega(\tau-\tau')}\mathcal{A}(\tau,\tau')\,. 
    \label{eq: transition-UDW}
\end{align}
The bi-distribution $\mathcal{A}(\tau,\tau')$ is the proper-time derivative of the Wightman two-point function $\mathsf{W}(\sx,\sx') = \braket{0|\hat\phi(\sx)\hat\phi(\sx')|0}$ pulled back along the detector's trajectory $\sx(\tau)$, i.e.,
\begin{align}
    \mathcal{A}(\tau,\tau') = \partial_\tau\partial_{\tau'}\mathsf{W}(\sx(\tau),\sx(\tau'))\,.
\end{align} 
Therefore, the detector response $P(\Omega)$ depends on the pullback of the Wightman function $\mathsf{W}(\sx(\tau),\sx(\tau'))$ along the detector's trajectory.

Finally, we need to specify the detector's trajectory. Since we will be comparing the Poincar\'e time machine with its universal cover, we need to restrict the detector's motion to be confined within the diamond-shaped region of the Poincar\'e patch with $\zeta_\pm>0$.  This is to ensure that the detector encounters no CTC anywhere during its interaction with the quantum field. Let us restrict our attention to the simple case of a  stationary trajectory $\xi=$ constant, namely
\begin{align}
    \sx(\tau) = (\eta(\tau),\xi) = (\W\xi\tau,\xi)\,,
    \label{eq: static-trajectory}
\end{align}
or in null coordinates $\zeta_\pm(\tau) = \xi\pm \W\xi\tau$. This trajectory has constant two-acceleration
\begin{align}
    a^\mu a_\mu = {\W}^2\,.
\end{align}
This acceleration is independent of $\xi$, which reflects the maximally symmetric nature of AdS$_2$. For the case of a time-machine geometry, in order to stay within the region without CTC we require that the detector-field interaction is confined to the region $\zeta_\pm >0$, i.e., the detector's interaction must be constrained to be within $|\eta(\tau)|\leq \xi$. More concretely, the requirement translates to 
\begin{align} 
    |\tau|\leq \W^{-1} = \frac{L}{\log A}\,. 
    \label{eq: tau-regime}
\end{align}
Thus the detector's interaction duration can be longer when the spacetime curvature is weaker.

\subsection{Calculation of the derivative two-point functions}

Recall that our goal is to analyse the detector response in the time machine geometry and understand the different regimes related to the strength of the time machine. For this, it will be very useful to evaluate the derivative two-point functions $\mathcal{A}(\tau,\tau')$ for Minkowski space, Einstein cylinder, and Poincar\'e-AdS$_2$ as they correspond to taking certain limits of the two-point function for the Poincaré time machine. Furthermore, these can be straightforwardly obtained from their well-known Wightman two-point functions, which in turn follow directly from standard mode-sum calculations and exploiting the conformal flatness of two-dimensional geometries for curved geometries (see, e.g., \cite{birrell1984quantum}). 

For convenience, let us first define the double-null coordinates $u=x-t,v=x+t$, so that by writing $\Delta t = t-t'$, $\Delta x = x-x$ we have $\Delta u =\Delta x - \Delta t$ and $\Delta v = \Delta x + \Delta t$. The Wightman two-point function for two-dimensional massless scalar field in Minkowski spacetime is well-known and is given by 
\begin{align}
    \mathsf{W}_{\textsc{M}}(\sx,\sx') &= -\frac{1}{4\pi}\log\rr{-\Lambda^2(-\Delta u-\ii\epsilon)(\Delta v-\ii\epsilon)}\,,
\end{align}
where $\Lambda>0$ is some IR cutoff. The dependence on the IR cutoff is the origin of the IR ambiguity for massless fields in two-dimensional Minkowski spacetime. The Wightman two-point function for the massless field in the Einstein cylinder is given by \cite{EMM2014zeromode}
\begin{align}
    \mathsf{W}_{\textsc{EC}}(\sx,\sx') 
    &= -\frac{1}{4\pi}\log\rr{1-e^{-\frac{2\pi \ii}{L}(-\Delta u - \ii\epsilon)}}\notag\\
    &\hspace{0.4cm} -\frac{1}{4\pi}\log\rr{1-e^{-\frac{2\pi \ii}{L}(\Delta v - \ii\epsilon)}}\,.
\end{align}
Here $L$ is the perimeter of the Einstein cylinder. Notice that the field is effectively given a periodic boundary condition so that $L$ serves as an IR cutoff. Hence, the IR ambiguity from Minkowski spacetime no longer appears in the cylindrical spacetime.

The Wightman two-point function for the massless field in the Poincar\'e-AdS$_2$ spacetime can be readily obtained from the mode functions in the Poincar\'e patch. Using the corresponding null coordinates in the Poincar\'e-AdS$_2$ patch $\zeta_\pm \coloneqq  \xi \pm \eta $, 
the Wightman two-point functions for field living in the Poncar\'e-AdS$_2$ patch is given by
\begin{align}
    \mathsf{W}_{\text{AdS}_2}(\sx,\sx')  &= -\frac{1}{4\pi}\log\left[\frac{(\Delta \zeta_+ - \ii\epsilon)(-\Delta \zeta_- - \ii\epsilon)}{(-\zeta_--\zeta_+'-\ii\epsilon)(\zeta_++\zeta_-'-\ii\epsilon)}\right]\,,
    \label{eq: Wightman-Poincare}
\end{align}
Note that this is functionally similar to the case of a static mirror geometry in flat spacetime since $\zeta_\pm$ takes the role of $v,u$ coordinates respectively \cite{cong2019entanglement}. This follows directly from the fact that we needed to put the boundary condition at the conformal boundary of the patch, effectively putting a mirror at the boundary. Therefore, the Wightman function also does not have any IR ambiguity.

The derivative two-point functions $\mathcal{A}(\tau,\tau')$ for the different background geometries can now be readily obtained by taking proper-time derivatives with respect to the static trajectory~\eqref{eq: static-trajectory}:
\begin{subequations}
\begin{align}
    \mathcal{A}_{\textsc{M}}(\tau,\tau') &= \frac{-1}{2\pi(\tau-\tau')^2}\,,
    \label{eq: Mink-derivative}\\
    \mathcal{A}_{\textsc{EC}}(\tau,\tau';\gamma) &= -\frac{\pi}{2L^2}\text{csc}^2\Bigr[\frac{\pi(\tau-\tau')}{L}\Bigr] + \frac{\gamma}{2L^2}\,,
    \label{eq: EC-derivative}\\
    \mathcal{A}_{\text{AdS}_2}(\tau,\tau') &= \frac{-1}{2\pi(\tau-\tau')^2}\frac{16-12\W^2(\tau-\tau')^2}{\bigr[4-\W^2(\tau-\tau')^2\bigr]^2}\,.
    \label{eq: poincare-derivative}
\end{align}
\end{subequations}
The parameter $\gamma$ defines the zero mode regularization in Einstein cylinder \cite{EMM2014zeromode,tjoa2019zeroresponse}. Recall that the parameter $\text{W}\geq 0$ is the inverse AdS radius, corresponding to the constant Ricci scalar $R=-2\text{W}^{2}$. Clearly, from Eq.~\eqref{eq: EC-derivative} the Einstein cylinder Wightman function can be decomposed into the oscillator and zero mode two-point functions:
\begin{align}
    \mathcal{A}_{\textsc{EC}}(\tau,\tau';\gamma) = \mathcal{A}_{\textsc{EC}}^{\text{osc}}(\tau,\tau') + \mathcal{A}_{\textsc{EC}}^{\text{zm}}(\gamma)\,.
    \label{eq: EC-derivative-splitting}
\end{align}
Note that all the time dependence is carried by  the oscillator contribution.

Finally, we are ready to calculate the derivative of the  two-point function for the time-machine geometry. {From the automorphic construction, it can be shown that the Wightman two-point function of the field on a multiply-connected spacetime is related to the corresponding two-point function on its universal covering space via an image sum \cite{BanachDowker1979,Banach1979mathissues} (see also \cite{Ana2021timemachine}). This gives\footnote{At the technical level, we can interchange the sum and the derivatives because the sum converges in the distributional sense; alternatively, since numerically we will truncate the summation into finite number of terms from $n=-N$ to $n=N$, finite sums do not have convergence issue.}} 
\begin{align}
    \mathcal{A}_{\textsc{TM}}(\tau,\tau') &= \sum_{n=-\infty}^\infty \partial_\tau\partial_{\tau'}\mathsf{W}_{\text{AdS}_2}\bigr(\sx(\tau),A^n\sx(\tau')\bigr)\,.
    \label{eq: image-sum-TM-wightman}
\end{align}
Numerically it is convenient to evaluate this sum as finite truncations of the sum which serves as an ultraviolet (UV) cutoff.

Before we proceed with the detector response calculations, it is worth pointing out certain useful limits. First, we can check that the derivative Wightman two-point functions have good IR behaviour: 
\begin{align}
    \lim_{\W^{-1} \to \infty} \mathcal{A}_{\text{AdS}_2} &= \mathcal{A}_{\textsc{M}}\,,\quad 
    \lim_{L\to \infty } \mathcal{A}_{\textsc{EC}} = \mathcal{A}_{\textsc{M}}\,.
\end{align}
This tells us that the zero mode contribution in the Einstein cylinder scenario vanishes in the large $L$ limit and we recover the Minkowski spacetime. Similarly, we recover the Minkowski spacetime case when we set the AdS radius of curvature $\W^{-1}$ to zero. Observe that the oscillator component of the Einstein cylinder Wightman function $\mathcal{A}^{\text{osc}}_{\textsc{EC}}$ can itself be written as an image sum:
\begin{align}
    \mathcal{A}_{\textsc{EC}}^{\text{osc}}(\tau,\tau') &= \sum_{n=-\infty}^{\infty}\mathcal{A}_{\textsc{M}}(\tau,\tau'+nL)\,,
\end{align}
where the $n\neq 0$ terms scales like $L^{-2}$; hence the image sum provides the required oscillatory corrections to the flat-space Wightman functions. Second, for the derivative Wightman function in the Einstein cylinder, it is possible to actually set the zero-mode regulator $\gamma$ to zero: the original Wightman function $\mathsf{W}_{\textsc{EC}}$ has a \textit{constant} zero-mode divergence originating from the zero mode variance $\braket{\hat{\phi}_{\textsc{zm}}(t)\hat{\phi}_{\textsc{zm}}(t')}\sim \gamma^{-1}+ \mathcal{O}(\gamma)$: the derivative coupling removes the problematic $\gamma^{-1}$ term.
 
One of the main goals of this work is to understand the limiting behaviour of the time machine geometry. To illustrate this problem, suppose we would like to know the weak-curvature regime $\W \approx 0^+$. Because $\W = \log(A)/L$, there are actually two ways to take the limit: 
\begin{enumerate}[label=(\roman*),leftmargin=*]
    \item  $A =  1+\delta$, where $0< \delta \ll 1$ for fixed $L>0$, or 
    \item $L\to\infty$ while keeping $A>1$ finite.
\end{enumerate}
Intuitively, we may expect that Case (i) approaches the Einstein cylinder in the limit which has zero curvature and finite $L$, while Case (ii) approaches the Poincar\'e-AdS$_2$ in the limit since it is an open spacetime with nonzero curvature. We will see that this indeed the case in the next section.

\section{Detector response}
\label{sec: detector-response}

For simplicity, in this work we consider a Gaussian switching
\begin{align}
    \chi(\tau) = e^{-\frac{\tau^2}{T^2}}\,,
\end{align}
where the switching width $T$ prescribes the effective duration of the interaction. This allows for exact computations in some cases (e.g., Minkowski spacetime and Einstein cylinder) and is numerically favourable. However, there is a small technical detail we need to deal with: recall from Eq.~\eqref{eq: tau-regime} that we require $|\tau|\leq \W^{-1}$ in order for the interaction to be confined within the no-CTC region. From a numerical viewpoint, this seems to be at odds with our choice of non-compact Gaussian switching, but we will argue later that this choice is a matter of practical convenience. For Gaussian switching, the requirement in Eq.~\eqref{eq: tau-regime} amounts to constraining the Gaussian width $T$ to be
\begin{align}
    w\coloneqq\W T \ll 1\,.
\end{align}
This sets the scale for the size of $\W$ in units of $T$. 
We will also use the dimensionless energy gap $ \omega \coloneqq \Omega T$ and the dimensionless spatial period 
$ \ell = L/T$. Furthermore, it is convenient to write $A=1+\delta$ for some $\delta>0$, so that
\begin{align}
    w = \frac{\log (1+\delta)}{\ell}\,.
\end{align}
Note that for the derivative coupling model, the coupling constant $\lambda$ is dimensionless in Eq.~\eqref{eq: derivative-coupling-Hamiltonian} since $\hat{\phi}$ has dimension zero in (1+1) dimensions.
The requirement that $w\ll 1$ can be achieved in two ways, borrowing the terminology from \cite{emparan2021holography}:
\begin{enumerate}[leftmargin=*,label=(\roman*)]
    \item \textbf{Slow time machine regime:} for fixed $\ell$, choose $\delta\ll 1$ such that $w\approx \delta/\ell \ll 1$. 

    \item \textbf{Fast time machine regime:} for fixed $\delta$ (not necessarily small), choose $\ell\gg 1$ such that $w\ll 1$. 
\end{enumerate}
The terminology is motivated by the fact that $w$ measures the amount of redshift in the identification $\zeta_\pm \sim A\zeta_\pm \equiv (1+w l)\zeta_\pm$ for a fixed dimensionless circumference $l$. 
Because the dimensionless curvature scale $w$ is required to be small, we can think of both regimes as being weakly curved over the duration of interaction $T$. Furthermore, it is possible to set $w$ to be equal for Case (i) and (ii), hence the differences in the detector response is purely topological in nature. This generalizes the result in \cite{smith2016topology}.

For Minkowski spacetime, we see by inspection that $\mathcal{A}_{\textsc{M}}$ is proportional to the pullback of the (3+1)-dimensional Wightman function in Minkowski spacetime to an inertial observer at rest:
\begin{align}
    \mathcal{A}_{\textsc{M}}(\tau,\tau') = 2\pi\mathsf{W}_{\textsc{M}}^{(3+1)}(\tau,\tau').
\end{align}
Since the right hand side can be computed using the well-known plane-wave expansion given by
\begin{align}
\mathsf{W}_{\textsc{M}}^{(3+1)}(\tau,\tau') = \lim_{\epsilon\to 0^+}\int\frac{\dd^3\bk}{2(2\pi)^3|\bk|}e^{-\ii|\bk|(\tau-\tau'-\ii\epsilon)}\,,
\end{align}

the detector response \eqref{eq: transition-UDW} can be written in terms of the Fourier transform of the switching function:
\begin{align}
    P_{{\textsc{M}}} 
    &= \lambda^2 \int_0^\infty \dd k\,\frac{k}{2\pi} \left|\tilde\chi({\omega/T}+k)\right|^2 \notag\\
    &= \frac{\lambda^2}{2}\left(e^{-\frac{1}{2} \omega ^2}-\sqrt{ \frac{\pi}{2} } \omega  \,\text{erfc}\left(\frac{\omega }{\sqrt{2}}\right)\right)\,,
\end{align}
where $k=|\bk|$.  For the Einstein cylinder, we can use the fact that we can split
\begin{align}
    \mathcal{A}_{\textsc{EC}}(\tau,\tau') &= \mathcal{A}_{\textsc{M}}(\tau,\tau') + \mathcal{A}_{\textsc{EC}}^{\text{reg}}(\tau,\tau')\,,
\end{align}
where $\mathcal{A}_{\textsc{EC}}^{\text{reg}} = \mathcal{A}_{\textsc{EC}}-\mathcal{A}_{\textsc{M}}$ is a regular well-behaved function. This splitting leads to simpler numerical calculations as it reduces the computation of the detector response to a one-dimensional (numerical) integration
\begin{subequations}
\begin{align}
    P_{{\textsc{EC}}} &=  P_{{\textsc{M}}} + P_{{\textsc{EC}}}^{\text{reg}}\,,\\
    P_{{\textsc{EC}}}^{\text{reg}} &=  \lambda^2\sqrt{\frac{\pi T^2}{2}}\int_{-\infty}^\infty\dd u\, e^{-\frac{u^2}{2T^2}}e^{-\ii{\omega u/T}}\mathcal{A}_{\textsc{EC}}^{\text{reg}}(u,0)\,.
\end{align}
\end{subequations}
For AdS$_2$ spacetime, we can also do this: 
\begin{subequations}
\begin{align}
    P_{\text{AdS}_2} &=  P_{{\textsc{M}}} + P_{\text{AdS}_2}^{\text{reg}}\,,\\
    P_{{\text{AdS}_2}}^{\text{reg}} &= \lambda^2\sqrt{\frac{\pi T^2}{2}}\int_{C(\epsilon)}\dd u\, e^{-\frac{u^2}{2T^2}}e^{-\ii{\omega u/T}}\mathcal{A}_{\text{AdS}_2}^{\text{reg}}(u,0)\,,
\end{align}
\end{subequations}
where the contour $C(\epsilon)$ is over the real axis $\R$ but deformed to the lower complex plane near the simple poles $u=\pm 2\W^{-1}$ of $\mathcal{A}_{{\text{AdS}_2}}^{\text{reg}}$.

Finally, the detector response for the time machine geometry is not easy to calculate because we need the image sum \eqref{eq: image-sum-TM-wightman} and we no longer have stationarity property $\A(\tau,\tau') = \A(\tau-\tau',0)$ that the previous three geometries possess. However, we can still evaluate the integral directly by breaking \eqref{eq: image-sum-TM-wightman} term-wise and calculate the integrals separately. That is, we write
\begin{subequations}
\begin{align}
    P_{\textsc{TM}} &= \sum_{n=-\infty}^{\infty} P^{(n)}_{\textsc{TM}}\,,\\
    P^{(n)}_{\textsc{TM}} &= \lambda^2\int\dd\tau\,\dd\tau'\,\chi(\tau)\chi(\tau')e^{-\ii\Omega(\tau-\tau')}\notag \\
    &\hspace{2cm}\times \partial_{\tau}\partial_{\tau'}\mathsf{W}_{\text{AdS}_2}(\sx(\tau),A^n\sx(\tau'))\,. 
\end{align}
\end{subequations}
In practice, we will need to truncate the summation over $n$ from $-N$ to $N$ for sufficiently large $N$ and one can verify the convergence numerically. We also compare this with another expression for the time machine Wightman function where the real part is in terms of Jacobi elliptic theta function \cite{Ana2021timemachine}, and in the parameter regimes where both are numerically stable, they give the same results.

Before we present our results, let us address the issue mentioned earlier that the requirement \eqref{eq: tau-regime}  seems to be at odds with the choice of non-compact Gaussian switching function. In practice, due to the strong support of the Gaussian we truncate numerically the Gaussian to finite duration while performing an appropriate contour integration for the calculation of the excitation probability. This truncation may appear unsatisfactory from a mathematical standpoint, since sharp truncation can lead to UV divergences. Suppose we cut the Gaussian so that it is only supported at $J\coloneqq [-5T/2,5T/2]$ with effective width $5T$. Then the truncated Gaussian is equivalent to applying an indicator function supported on $J$, i.e., 
\begin{align}
    \chi_{J}(\tau) &\coloneqq \openone_{J}e^{-\frac{\tau^2}{T^2}} \notag\\
    \openone_J &\equiv  \Theta\rr{\tau+\frac{5T}{2}}- \Theta\rr{\tau-\frac{5T}{2}}
\end{align}
where $\Theta(z)$ is the Heaviside function. UV divergences can arise because of the distributional singularities of the Wightman function coinciding with the discontinuity of the Heaviside function. It is possible to verify numerically that if we add a UV regulator $\epsilon$ on the Heaviside function by 
\begin{align}
    \Theta_\epsilon(z)\coloneqq \frac{1}{2}(1+\tanh(z/\epsilon))\,,
\end{align}
the excitation probability for sufficiently small $\epsilon$ is indistinguishable from using the full Gaussian switching. In effect what the UV regulator does is to ``smoothen the corners'' in $\chi_J(\tau)$, and from this we can instead interpret the original Gaussian switching as ``UV-regulated'' version of compact, truncated switching demanded by Eq.~\eqref{eq: tau-regime}. One could also insist on using genuine smooth compactly supported switching functions as in \cite{louko2006often,satz2007then}, but since the UV-divergent piece is purely a property of the switching and not about the trajectory, background spacetime or the detector parameters, this will not modify our analysis in any significant manner.

\begin{figure}[tp]
    \centering
    \includegraphics[scale=0.9]{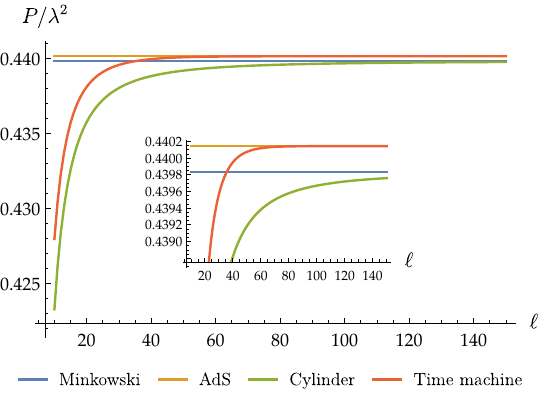}
    \caption{Detector response as a function of dimensionless circumference $\ell=\frac{\log A}{w}$. We fix the dimensionless parameters to be $\omega=0.1,w=0.05,\gamma=0.01,N=10$. In the limit of large $\ell$, the detector response for the time machine geometry approaches the value of the detector response in the Poincar\'e-AdS$_2$ patch. In the limit of small $\ell$, the detector response approaches instead to the case of Einstein cylinder.}
    \label{fig: derivative-coupling-0}
\end{figure}

Let us first demonstrate the transition from slow to fast time machine regimes by fixing the dimensionless curvature scale $w=1/20$, and vary $\ell\in [10,150]$. For clarity, we chose to vary $\ell=\frac{\log A}{w}$ to change the strength of the time machine since $\ell$ for fixed curvature $w$ is equivalent to varying $A$.

Figure~\ref{fig: derivative-coupling-0} shows that in the limit of large $\ell$, corresponding to the fast time-machine, the detector response for the time machine geometry approaches the value of the detector response in the Poincar\'e-AdS$_2$ patch. Since the local curvature scale is fixed by $w$, this shows that the fast time machine geometry is locally indistinguishable from the standard AdS background. In contrast, if the time machine geometry is slow (relative to interaction time $T$), then the detector response becomes more similar to the detector response in the Einstein cylinder. The deviation always persists since the Einstein cylinder has zero spacetime curvature while the time machine geometry is set to have finite nonzero curvature. Note that we have fixed the local curvature parameter $w$ of the time machine geometry to be equal to the corresponding Poincar\'e-AdS$_2$ limit: consequently, the differences in the detector responses as we vary $\ell$ is purely a topological in nature. This can be regarded as a curved spacetime generalization to the result in \cite{smith2016topology} that distinguishes Minkowski spacetime from Einstein cylinder spacetime in (3+1) dimensions\footnote{However, there is no zero-mode contribution because only one spatial direction is identified.}.

\begin{figure}[tp]
    \centering
    \includegraphics[scale=0.9]{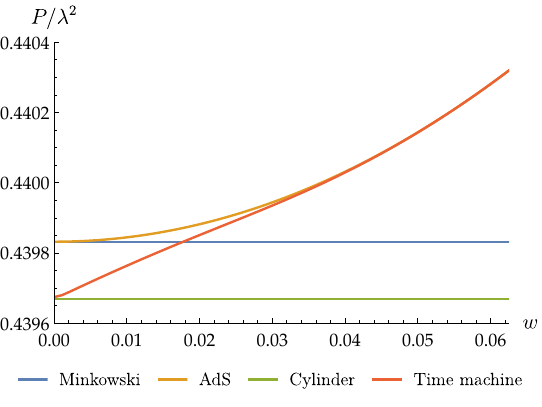}
    \caption{Detector response as a function of dimensionless curvature parameter $w$. We fix the dimensionless parameters to be $\omega=0.1,\ell=100,\gamma=0.01,N=15$. In the limit of large $w$, the detector response for the time machine geometry approaches the value of the detector response in the Poincar\'e-AdS$_2$ patch. In the limit of small $w$, the detector response approaches instead to the case of Einstein cylinder.} 
    \label{fig: derivative-coupling-1}
\end{figure}

A different way of interpreting the slow and fast time machine regimes can be given if we now fix $\ell$ and find how the detector response varies as a function of curvature parameter $w$. We chose $\ell=100$ to ensure that the spacetime is large enough for the support of the switching to be far from the Cauchy horizons. The results are shown in Figure~\ref{fig: derivative-coupling-1}. In the large curvature regime, the detector response for the time machine geometry approaches the value of the detector response in the Poincar\'e-AdS$_2$ patch, while for weak curvature the detector response approaches that of the Einstein cylinder. An equivalent way of saying this is that if the time machine geometry is chosen to be of the same ``size'' as the Einstein cylinder, the strength of the time machine is controlled by curvature. Figure~\ref{fig: derivative-coupling-1} also seems to suggest another interesting interpretation: since Poincar\'e-AdS$_2$ is an open universe, it appears as though stronger gravitational fields make it harder to distinguish the global topology of the spacetime, while for very small $w$ the detector can distinguish Minkowski spacetime from Einstein cylinder through the constant zero-mode contribution. 

\section{Conclusion and outlook}

In this work we have shown that local measurements carried out with particle detectors can reveal if the spacetime where the detector moves possess CTCs. Remarkably, this is true even though the CTCs are causally disconnected from the detectors and protected by a horizon. We have done this by studying a particular spacetime with CTCs (arising from the topological identification of the Poincar\'e patch of two-dimensional AdS spacetime) and comparing the limits where the time warp in the CTCs is strong and when the limit where the CTCs disappear (and we recover the flat Einstein cylinder). We have also compared the response of the detector in the time machine case with the response of the detector in pure AdS, showing that the detector can distinguish between effects of the geometry of spacetime and the topological effects associated with the spacetime's chronological structure. 

In \cite{emparan2021holography} the holographic bulk dual to a time machine geometry is constructed in order to study the connection between chronology protection and the geometry of the bulk spacetime dual to the time machine geometry. One crucial difference 
is that the field that lives in the time machine geometry is taken to be a strongly interacting conformal field theory that admits an explicit semiclassical dual spacetime. It is unclear whether two time machine regimes, which is separated by the size of the zero-mode phenomenon in the free theory, has an analog in the strongly interacting case and if it does, what is the corresponding statement in the bulk dual. The study of UDW model coupled to strongly-interacting fields has also been largely unexplored. We leave these two questions for future work.

\acknowledgments

The authors thank Jorma Louko for discussion and useful comments.
A. A.-S. is funded by the Deutsche Forschungsgemeinschaft (DFG, German Research
Foundation) — Project ID 516730869. E. T. acknowledges funding from the Munich Center for Quantum Science and Technology (MCQST), funded by the Deutsche Forschungsgemeinschaft (DFG) under Germany’s Excellence Strategy (EXC2111 - 390814868).  This work was also partially supported by Spanish Project No. MICINN PID2020-118159GB-C44. E. M.-M. acknowledges support through the Discovery Grant Program of the Natural Sciences and Engineering Research Council of Canada (NSERC).

\bibliography{ref}

\providecommand{\href}[2]{#2}\begingroup\raggedright\begin{thebibliography}{10}

\bibitem{BanachDowker1979}
R.~Banach and J.~S. Dowker, {\it The vacuum stress tensor for automorphic fields on some flat space-times},  {\em J. Phys. A} {\bf 12} (dec, 1979) 2545--2562.

\bibitem{Banach1980automorphic}
R.~Banach, {\it The quantum theory of free automorphic fields},  {\em J. Phys. A} {\bf 13} (jun, 1980) 2179--2203.

\bibitem{Frolov1991locallystatic}
V.~P. Frolov, {\it Vacuum polarization in a locally static multiply connected spacetime and a time-machine problem},  {\em Phys. Rev. D} {\bf 43} (Jun, 1991) 3878--3894.

\bibitem{EMM2014zeromode}
E.~Mart{\'\i}n-Mart{\'\i}nez and J.~Louko, {\it Particle detectors and the zero mode of a quantum field},  {\em Phys. Rev. D} {\bf 90} (2014), no.~2 024015.

\bibitem{tjoa2019zeroresponse}
E.~Tjoa and E.~Mart{\'\i}n-Mart{\'\i}nez, {\it Zero mode suppression of superluminal signals in light-matter interactions},  {\em Phys. Rev. D} {\bf 99} (2019), no.~6 065005.

\bibitem{tjoa2020harvesting}
E.~Tjoa and R.~B. Mann, {\it Harvesting correlations in schwarzschild and collapsing shell spacetimes},  {\em J. High Energy Phys.} {\bf 2020} (Aug, 2020).

\bibitem{Page2012deSitter}
D.~N. Page and X.~Wu, {\it Massless scalar field vacuum in de sitter spacetime},  {\em J. Cosmol. Astropart. Phys.} {\bf 2012} (nov, 2012) 051--051.

\bibitem{Allen1987deSitter}
B.~Allen and A.~Folacci, {\it Massless minimally coupled scalar field in de sitter space},  {\em Phys. Rev. D} {\bf 35} (Jun, 1987) 3771--3778.

\bibitem{witten2021does}
E.~Witten, 2021.

\bibitem{Isham1978}
S.~J. Avis, C.~J. Isham, and D.~Storey, {\it Quantum field theory in anti-de sitter space-time},  {\em Phys. Rev. D} {\bf 18} (Nov, 1978) 3565--3576.

\bibitem{wald1980nonhyperbolic}
R.~M. Wald, {\it Dynamics in nonglobally hyperbolic, static space‐times},  {\em J. Math. Phys.} {\bf 21} (1980), no.~12 2802--2805.

\bibitem{Ishibashi2004AdS}
A.~Ishibashi and R.~M. Wald, {\it Dynamics in non-globally-hyperbolic static spacetimes: {III}. anti-de sitter spacetime},  {\em Class. Quantum Gravity} {\bf 21} (may, 2004) 2981--3013.

\bibitem{Dappiaggi2016PAdS}
C.~Dappiaggi and H.~R.~C. Ferreira, {\it {Hadamard states for a scalar field in anti--de Sitter spacetime with arbitrary boundary conditions}},  {\em Phys. Rev. D} {\bf 94} (Dec, 2016) 125016.

\bibitem{Dappiaggi2018CAdS}
C.~Dappiaggi, H.~R.~C. Ferreira, and A.~Marta, {\it {Ground states of a Klein-Gordon field with Robin boundary conditions in global anti--de Sitter spacetime}},  {\em Phys. Rev. D} {\bf 98} (Jul, 2018) 025005.

\bibitem{maldacena1999largeN}
J.~Maldacena, {\it The large-n limit of superconformal field theories and supergravity},  {\em International journal of theoretical physics} {\bf 38} (1999), no.~4 1113--1133.

\bibitem{Witten:1998zw}
E.~Witten, {\it {Anti-de Sitter space, thermal phase transition, and confinement in gauge theories}},  {\em Adv. Theor. Math. Phys.} {\bf 2} (1998) 505--532.

\bibitem{de2001holographic}
S.~de~Haro, K.~Skenderis, and S.~N. Solodukhin, {\it Holographic reconstruction of spacetime and renormalization in the ads/cft correspondence},  {\em Commun. Math. Phys.} {\bf 217} (2001), no.~3 595--622.

\bibitem{Ana2021timemachine}
A.~Alonso-Serrano, E.~Tjoa, L.~J. Garay, and E.~Martín-Martínez, {\it The time traveler’s guide to the quantization of zero modes},  {\em J. High Energy Phys.} {\bf 2021} (Dec, 2021) 170.

\bibitem{emparan2021holography}
R.~Emparan and M.~Toma{\v{s}}evi{\'c}, {\it Holography of time machines},  {\em J. High Energy Phys.} {\bf 2022} (2022), no.~3 1--33.

\bibitem{Unruh1979evaporation}
W.~G. Unruh, {\it Notes on black-hole evaporation},  {\em Phys. Rev. D} {\bf 14} (Aug, 1976) 870--892.

\bibitem{DeWitt1979}
B.~S. {Dewitt}, {\it {Quantum gravity: the new synthesis}},  in {\em General Relativity: An Einstein centenary survey} (S.~W. {Hawking} and W.~{Israel}, eds.), pp.~680--745, 1979.

\bibitem{Lopp2021deloc}
R.~Lopp and E.~Mart\'{\i}n-Mart\'{\i}nez, {\it Quantum delocalization, gauge, and quantum optics: Light-matter interaction in relativistic quantum information},  {\em Phys. Rev. A} {\bf 103} (Jan, 2021) 013703.

\bibitem{Aubry2014derivative}
B.~A. Ju{\'{a}}rez-Aubry and J.~Louko, {\it Onset and decay of the 1 + 1 hawking-unruh effect: what the derivative-coupling detector saw},  {\em Class. Quantum Gravity} {\bf 31} (nov, 2014) 245007.

\bibitem{juarez2018quantum}
B.~A. Ju{\'a}rez-Aubry and J.~Louko, {\it Quantum fields during black hole formation: how good an approximation is the unruh state?},  {\em J. High Energy Phys.} {\bf 2018} (2018), no.~5 1--24.

\bibitem{tjoa2022unruhdewitt}
E.~Tjoa and R.~B. Mann, {\it Unruh-dewitt detector in dimensionally-reduced static spherically symmetric spacetimes},  {\em J. High Energy Phys.} {\bf 2022} (2022), no.~3 1--30.

\bibitem{Deser1997}
S.~Deser and O.~Levin, {\it Accelerated detectors and temperature in (anti-) de sitter spaces},  {\em Class. Quantum Gravity} {\bf 14} (sep, 1997) L163--L168.

\bibitem{Jennings2010}
D.~Jennings, {\it On the response of a particle detector in anti-de sitter spacetime},  {\em Class. Quantum Gravity} {\bf 27} (sep, 2010) 205005.

\bibitem{Hodgkinson2012BTZ}
L.~Hodgkinson and J.~Louko, {\it Static, stationary, and inertial unruh-dewitt detectors on the btz black hole},  {\em Phys. Rev. D} {\bf 86} (Sep, 2012) 064031.

\bibitem{Henderson2020antihawking}
L.~J. Henderson, R.~A. Hennigar, R.~B. Mann, A.~R. Smith, and J.~Zhang, {\it Anti-hawking phenomena},  {\em Phys. Lett. B} {\bf 809} (2020) 135732.

\bibitem{Pitelli2021ads2}
J.~P.~M. Pitelli, B.~S. Felipe, and R.~A. Mosna, {\it Unruh-dewitt detector in ${\mathrm{ads}}_{2}$},  {\em Phys. Rev. D} {\bf 104} (Aug, 2021) 045008.

\bibitem{smith2016topology}
E.~Mart\'{\i}n-Mart\'{\i}nez, A.~R.~H. Smith, and D.~R. Terno, {\it Spacetime structure and vacuum entanglement},  {\em Phys. Rev. D} {\bf 93} (Feb, 2016) 044001.

\bibitem{KeithCo}
K.~K. Ng, R.~B. Mann, and E.~Mart\'{\i}n-Mart\'{\i}nez, {\it Equivalence principle and qft: Can a particle detector tell if we live inside a hollow shell?},  {\em Phys. Rev. D} {\bf 94} (Nov, 2016) 104041.

\bibitem{Aida1}
A.~Ahmadzadegan, E.~Mart\'{\i}n-Mart\'{\i}nez, and R.~B. Mann, {\it Cavities in curved spacetimes: The response of particle detectors},  {\em Phys. Rev. D} {\bf 89} (Jan, 2014) 024013.

\bibitem{KeithCo2}
K.~K. Ng, R.~B. Mann, and E.~Mart\'{\i}n-Mart\'{\i}nez, {\it Over the horizon: Distinguishing the schwarzschild spacetime and the $\mathbb{R}{\mathbb{p}}^{3}$ spacetime using an unruh-dewitt detector},  {\em Phys. Rev. D} {\bf 96} (Oct, 2017) 085004.

\bibitem{Smith_CQG}
A.~R.~H. Smith and R.~B. Mann, {\it Looking inside a black hole},  {\em Class. Quantum Gravity} {\bf 31} (apr, 2014) 082001.

\bibitem{Aida2}
A.~Ahmadzadegan, F.~Lalegani, A.~Kempf, and R.~B. Mann, {\it Probing geometric information using the unruh effect in the vacuum},  {\em Phys. Rev. D} {\bf 100} (Oct, 2019) 085013.

\bibitem{MachineLearning}
D.~Grimmer, I.~Melgarejo-Lermas, J.~Polo-G{\'o}mez, and E.~Mart{\'i}n-Mart{\'i}nez, {\it Decoding quantum field theory with machine learning},  {\em J. High Energy Phys.} {\bf 2023} (Aug, 2023) 31.

\bibitem{Dowker1972multiplyconnected}
J.~S. Dowker, {\it Quantum mechanics and field theory on multiply connected and on homogeneous spaces},  {\em J. Phys. A} {\bf 5} (jul, 1972) 936--943.

\bibitem{Banach1979mathissues}
R.~Banach and J.~S. Dowker, {\it Automorphic field theory-some mathematical issues},  {\em J. Phys. A} {\bf 12} (dec, 1979) 2527--2543.

\bibitem{birrell1984quantum}
N.~D. Birrell and P.~Davies, {\em Quantum Fields in Curved Space}.
\newblock Cambridge Monographs on Mathematical Physics. Cambridge University Press, 1984.

\bibitem{cong2019entanglement}
W.~Cong, E.~Tjoa, and R.~B. Mann, {\it Entanglement harvesting with moving mirrors},  {\em J. High Energ. Phys} {\bf 2019} (2019), no.~6 21.

\bibitem{louko2006often}
J.~Louko and A.~Satz, {\it How often does the unruh--dewitt detector click? regularization by a spatial profile},  {\em Class. Quantum Gravity} {\bf 23} (2006), no.~22 6321.

\bibitem{satz2007then}
A.~Satz, {\it Then again, how often does the unruh--dewitt detector click if we switch it carefully?},  {\em Class. Quantum Gravity} {\bf 24} (2007), no.~7 1719.

\end{thebibliography}\endgroup
\bibliographystyle{JHEP}

\end{document}